\documentclass[pr, twocolumn, preprintnumbers, showpacs, footnoteadded, superscriptaddress,longbibliography]{revtex4-1}
\usepackage[T1]{fontenc}
\usepackage{amssymb}
\usepackage{graphicx}
\usepackage{amsmath}
\usepackage{slashed}
\usepackage{tensor}
\usepackage{hyperref}
\usepackage{epstopdf}
\usepackage{extarrows}
\usepackage{xcolor}
\usepackage{epstopdf}
\usepackage{tikz}
\usepackage{ulem}

\newcommand{\td}{\text{d}}
\newcommand{\te}{\text{e}}
\newcommand{\ti}{\text{i}}

\newcommand{\be}{\begin{equation}}
\newcommand{\ee}{\end{equation}}
\newcommand{\bea}{\setlength\arraycolsep{2pt} \begin{eqnarray}}
\newcommand{\eea}{\end{eqnarray}}

\newcommand{\bpm}{\begin{pmatrix}}
\newcommand{\epm}{\end{pmatrix}}

\newcommand{\ZF}[1]{\textcolor{red}{#1}}

\begin{document}
%\title{Inherent Instabilities in Stable Black Holes}
\title{Numerical study on the robustness of the stability for stable black holes}
\author{Zhan-Feng Mai}
%\email{zhanfeng.mai@gmail.com}
\affiliation{Guangxi Key Laboratory for Relativistic Astrophysics, School of Physical Science and Technology, Guangxi University, Nanning 530004, P.~R.~China}
\author{Run-Qiu Yang}
\email{aqiu@tju.edu.cn}
\affiliation{Center for Joint Quantum Studies and Department of Physics, School of Science, Tianjin University, Yaguan Road 135, Jinnan District, 300350 Tianjin, P.~R.~China}

\begin{abstract}
This paper numerically studies if the stability of a stable black hole is robust against the small perturbation on geometry near its event horizon. In an other word, we numerically study if two nearly identical black holes may exhibit completely different stabilities at late time. As a toy model, it encodes the such perturbation into deformations of Regge-Wheeler potential. It considers three different types of local deformations-the negative static bump potential, the stochastic potential and bump potential modulated by time function in low frequency limit. Our numerical results show that infinitesimal local deformations on Regge-Wheeler potential near the horizon can overturn stability of a stable black hole, implying that late-time behavior of a stable black hole is extremely sensitive to geometry near horizon. Specially, certain deformations that stabilize systems in flat backgrounds can destabilize otherwise stable black holes. It also shows that horizon-induced redshift transforms near-horizon quantum fluctuations into classical-scale stochastic deformations capable of triggering instability, implying that even an isolated black hole cannot keep stable if the near-horizon quantum noise could be hold in extended timescales.
\end{abstract}

%\end{abstract}
%%%%%%%%%%%%%%%%%%%%%%%%%%%%%%%%%%%%%%
\maketitle
%\tableofcontents
\flushbottom

%%%%%%%%%%%%%%%%%%%%%%%%%%%%%%%%%%%%%%
\noindent

\section{Introduction}
In the field of black hole (BH) physics, the analysis of BH stability stands as a fundamental and critical subject, as it reveals whether BHs can maintain their existence in the universe over extended periods. The stability of a BH addresses the question: if it is subjected to a small perturbation, will the perturbations decay with time, allowing the BH to return to its original state, or will they grow uncontrollably and destroy the configuration altogether? The stability problem has deep implications for the mathematical structure of general relativity, cosmic censorship, and the no-hair conjecture.\cite{chandrasekhar1983mathematical}. The quasi-normal modes (QNMs), meanwhile, serve as one of the essential physical quantities for exploring and understanding the stability characteristics of BHs\cite{Chandrasekhar:1975zza,Nollert:1998ys,Kokkotas:1999bd,Berti:2009kk,Konoplya:2011qq}. Unlike normal modes in closed system, the QNMs are a series of discrete complex modes. The sign of the imaginary parts indicates whether the spacetime is stable or not. In general, QNMs indicating unstable spacetime describe a system where phase transition happens, such as spontaneous scalarization of compact object\cite{Silva:2017uqg,Herdeiro:2018wub,Doneva:2022ewd,Xu:2024cfe,Xiong:2024urw}, while QNMs describing stable spacetime play an important role in ringdown signal of gravitational waves in the last period of two astrophysical compact object merger\cite{Creighton:1999pm,Dreyer:2003bv,Berti:2005ys,Gossan:2011ha,Berti2018,Wang2024}.

Recently, an interesting feature of QNMs, even the fundamental modes, has been reported that they are \emph{sensitive} to small alterations of background, which is called spectrum instability\cite{Nollert:1996rf,Cheung:2021bol,Cardoso:2024mrw,Qian:2024iaq}. For example, when the effective potential is altered by adding a small positive ``bump'' away from the location of photon ring, the quasi-normal modes (QNMs) of a Schwarzschild BH are completely altered\cite{Cheung:2021bol}.In addition, the spiral formula of spectrum instability can be semi-analytically derived in double rectangular barriers using perturbation theory (See the Supplemental material in \cite{Motohashi:2024fwt}) The sensitivity of QNMs seems to imply that the QNMs are not appropriate observables in ringdown stage. However, the ringdown signal appears to be robust by the spectral instability~\cite{Daghigh:2020jyk,Kyutoku:2022gbr,Berti:2022xfj,Yang:2024vor,Ianniccari:2024ysv,Rosato:2024arw}. Specifically, it was pointed out that the spectrum instability can hardly affect the greybody factor and on the ringdown signal. Further, the phase shift of the greybody factor could carry more physical information when studying the environmental effects on QNMs and the ringdown signal~\cite{Kyutoku:2022gbr}.

During the early ringdown stage, the signal in altered background exhibits no substantial divergence from the original signal. In the late ringdown stage, while a discernible discrepancy arises between the two, the magnitude of this difference is directly proportional to the intensity of the bump and will exponentially decays to zero. Notably, both signals undergo rapid decay to zero, maintaining consistency in their asymptotic behavior. The ringdown signal is insensitive against the small positive alterations of effective potential. From a more macroscopic perspective, the stability of BHs remains unchallenged by this minor alteration on effective potential.
%The ``spectrum instability'' in those studies does not cause the instability of BHs in time domain.

When addressing the topic of BH stability, the focus is linear stability of BHs, which usually focuses on whether a probe field exhibits exponential growth in a fixed spacetime background. Linear stability analysis of BHs involves solving wave-like equations for perturbations of the metric or matter fields \cite{chandrasekhar1983mathematical}. Numerous studies have demonstrated that Schwarzschild BH remains stable under perturbations from scalar fields, Dirac fields, and spin-2 fields\cite{Regge:1957td,mtbh}. Due to the complex environment and unavoidable classical/quantum fluctuations of real BHs, no models in these studies can fully characterize any real BH in the universe. However, these investigations are still regarded as highly valuable. A key reason is the prevailing belief that the conclusion about BH stability is robust against small alterations of the spacetime metric-that is, sufficiently small alteration on the metric would not disrupt the stability of an initially stable BH. In other words, if a theoretical BH described by metric $g_{\mu\nu}$ is stable, then we believe the real BH described by metric $\tilde{g}_{\mu\nu}$ will also be stable if the difference between $\tilde{g}_{\mu\nu}$ and $g_{\mu\nu}$ is ``small'' enough. This in fact implies a hypothesis: \\
\indent\textit{$\bullet$ The stability of a stable black hole is robust against the small deformation on the geometry. }\\
This naive idea is seemingly supported by some models in studying ``spectrum instability''~\cite{Destounis2024}, since the perturbed potentials in those models, though cause drastic changes of the spectrum, do not change instability of BHs in time domain.

However, our results in this work suggest that this ``taken-for-granted'' assumption may not be valid. Using the Schwarzschild BH as a concrete example, we consider if its stability is sensitive to the modification of geometry near horizon. As a toy model, we encode such modification into the alteration of Regge-Wheeler potential. We analyze the long-time behavior of probe scalar field when a small negative or zero-mean random local effective potential is added sufficiently close to the horizon, extending the analysis beyond the positive bump scenario investigated in\cite{Cheung:2021bol}. In the time domain, we numerically show that the spacetime becomes unstable when the alteration is sufficiently close to the horizon. Particularly, certain perturbed potentials that stabilize systems in flat backgrounds can destabilize otherwise stable BHs. We conclude that, at least under static spherical symmetry, BH stability does not exhibit robustness.  Two nearly identical BHs may exhibit completely different stability properties at late time. It will also show that horizon-induced redshift transforms near-horizon quantum fluctuations into classical-scale stochastic local deformations on the metric, which can trigger instability of the BH and cause it gradually to lose energy over time. This implies that even an isolated black hole cannot keep stable if the near-horizon quantum noise could be hold in extended timescales.
In this paper, we use natural units $\hbar = c = G = 1$.

\section{Effective potential and its alterations}
We take the original spacetime to be Schwarzschild BH, of which the line element is
\begin{eqnarray}\label{dsmetric}
&& \td s^2 = -h \td t^2 + \frac{\td r^2}{h} + r^2 \td\Omega^2 \, , \cr
&& h = 1-\frac{r_h}{r} \, ,
\end{eqnarray}
where $r_h$ is the radius of event horizon, $\td\Omega^2$ stands for line element of unit sphere. In time domain, we then consider equation of motion with respect to the  s-wave scalar perturbation $\Psi=\psi(t,r_*)/r$  \emph{as a test field} in the Schwarzschild spacetime
\begin{equation}\label{maseq}
\frac{\partial^2 \psi}{\partial t^2 } - \frac{\partial ^2 \psi}{\partial r_*^2} + V_{\mathrm{eff}} \psi = 0\,.
\end{equation}
The tortoise coordinate $r_*$ is defined as $\td r_*/ \td r = 1/h$, implying that $r \to r_h, r_* \to -\infty$ and $r \to \infty, r_* \to \infty$. The effective potential in Eq.~\eqref{maseq} is just the Regge-Wheeler potential $V_{\mathrm{eff}} =V_{\mathrm{RW}}= h h' /r$. The $V_{\mathrm{RW}}$ is nonnegative, so the Schwarzschild BH is stable against the scalar perturbation, which has been confirmed in\cite{Regge:1957td}.

We now add an alteration term into the effective potential to capture possible environmental effects presenting around the BH system. Detailed discussion on how the environments change Eq.~\eqref{maseq} would be much involved and we here simply assume that they are all encoded into an alteration on effective potential
\begin{equation}\label{pertV1}
  V_{\mathrm{eff}}=V_{\mathrm{RW}}(r_*)+\epsilon V_p(r_*-a,t)\,,
\end{equation}
where $V_p(x,t)$ is a function that localizes around at $x=0$ and decays in spatial direction at least as fast as $V_{\mathrm{RW}}$. Note that $V_p(x,t)$ may not decay in time. \emph{Though it might not characterize details of any realistic setup and the dynamical evolution of matter and gravity might also need to be considered, it is usually used as a toy model and a beginning point in understanding how environments change the stability of the BH}~\cite{Cheung:2021bol,PhysRevD.89.104059,PhysRevD.104.084028,Capuano:2024qhv}.

In the following discussions, we will specify $r_h=1$. We present our numerical results for solving Eq.~\eqref{maseq} in time domain with initial data corresponding to Gaussian wave packet

\begin{equation}
\psi(r_*,0) = \exp\left(-\frac{r_*^2}{20}\right)\, , \quad \frac{\partial \psi}{\partial t}(r_*,0) = 0 \, .
\end{equation}

For spatial grid $r_*$, we adopt fourth-order central difference method for spatial derivative and integrate the Eq.~\eqref{maseq} using fourth-order Runge-Kutta method for $t$. At the horizon ($r_*\rightarrow-\infty$) and the infinity  ($r_*\rightarrow\infty$) we impose the no reflecting boundary conditions. We give further detailed description in Appendix \ref{numer}. In following we will consider three different types of potential $V_p(r_*-a,t)$.

\section{Static negative bump}
The first one is the ``bump'' static potential $V_p(x,t)=V_p(x)$, where $V_p(x)$ has certain sign and compact support. When $V_p(x)$ is positive, it has been well known that such alteration will lead to ``spectrum instability''. However, this kind of ``instability'' is not the instability of BH itself, i.e., the system will still settle down into a static BH without scalar hair. Since the bumps are added by hand, there is no proper justification for a positive or negative sign. Particularly, near the event horizon, the quantum effects of gravity and matter may lead to exotic matter and may raise negative ``bump''. The negative ``bump'' can even appear if we impose dominant energy condition (see one example in Appendix \ref{nebump}). The negative "bump" could resemble a tachyonic field or be inherent in various modified gravity theories, such as the Einstein-Gauss-Bonnet theory. This feature might give rise to spacetime instabilities~\cite{PhysRevD.80.127502,PhysRevD.108.104020,Cao:2024sot}. If the bump is close to a stable BH, the situation becomes complicated. We expect a competition with the absorption at the horizon. Naively, one may expect that the BH will be still stable if the original BH is strongly enough but the negative ``bump'' is small enough. Surprisingly, our numerical results show that, if the infinitesimal negative ``bump'' is close to event horizon enough, it can overturn an arbitrarily stable BH.

We specify the negative bump to be Gaussian form around center $r_*=a$
\begin{equation}\label{guassian1}
  V_p(r_*-a)=-\exp\left[-\frac{(r_*-a)^2}{16}\right]\,.
\end{equation}
We numerically investigate how this negative bump changes the long-time behavior of scalar perturbation when it is placed close to even horizon.

\begin{figure}[hbpt]
  \centering
  \includegraphics[width=0.5\textwidth]{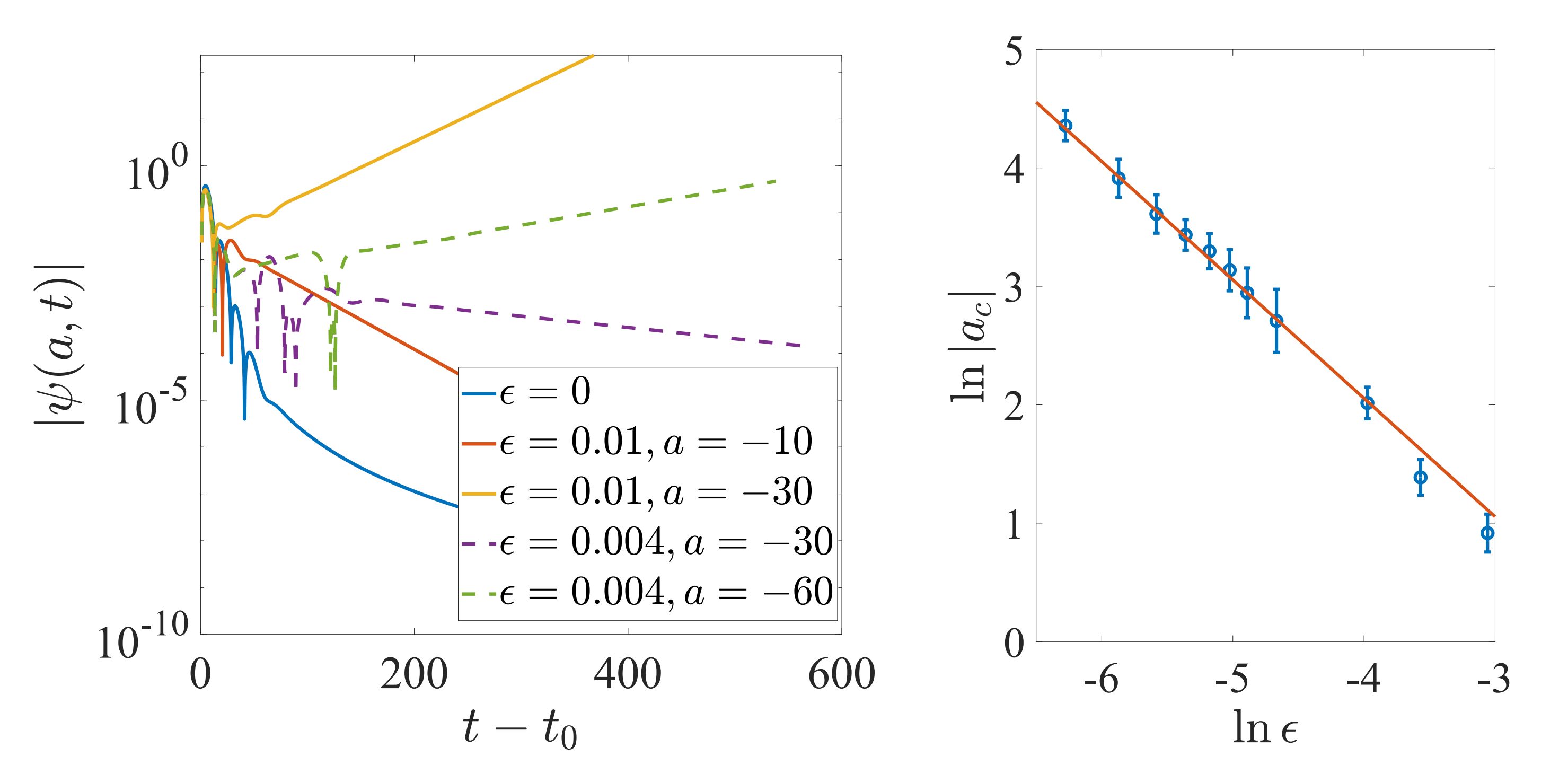}
  \caption{Left panel: Time evolutions of $\psi$ at $r_*=a$ for different values of $\epsilon$ and $a$. Here $t_0$ is defined as the first time at which $\psi(a,t)$ is at its local maximum value. Right panel: Relationship between $\epsilon$ and $|a_c|$. The error bars are determined by the bounds of bisection. The red line is given by equation $\ln|a_c|=-\ln|\overline{V}_p|-\ln\epsilon$ (not by fitting). }\label{bumpfig1}
\end{figure}

The left panel of Fig.~\ref{bumpfig1}.  shows the results as we move perturbed potential towards event horizon by choosing $\epsilon = 10^{-2}$ and $4\times10^{-3}$. It shows that for small values of $a$ where the alterations are confined near the photon ring, $\psi$ will decay to approach zero in the last stage, confirming that the Schwarzschild BH is stable. However, for the cases $\epsilon=10^{-2}$ we found that there exists a critical distance $a_c \approx -15$ of alteration to trigger the BH instability. When $a<a_c$, the perturbation field $\psi$ will experience exponential growth at late time. This exponential amplification signals an onset of instability, indicating that the perturbation field will drive the Schwarzschild BH towards an unstable state during the later stage of its evolution.

If we decrease the amplitude $\epsilon$, then we find that the critical distance will also move towards horizon. Using the method of bisection, we can approximately find the relationship between $\epsilon$ and $a_c$, which is shown in the right panel of Fig.~\ref{bumpfig1}. Particularly, we find following asymptotically relation between $\epsilon$ and $a_c$
\begin{equation}\label{realepac1}
  a_c=1/(\overline{V}_p\epsilon)\,
\end{equation}
when $\epsilon$ is small enough. Here $\overline{V}_p=\int_{-\infty}^{\infty}V_p(x)\td x$. We have tested various different types of negative bumps and find that Eq.~\eqref{realepac1} is true in all cases. Our numerical results imply that the instability will always be triggered if the negative bump is sufficiently close to event horizon no matter how small the amplitude is.

Besides, W. L. Qian et al. studied a similar configuration of effective potential in an asymmetric Damour-Solodukhin wormhole, which consists of two distinct black hole effective potentials separated by a distance~\cite{Qian:2025occ}. They considered the case of asymptotic echo modes $\omega \to \infty$ and found that the imaginary part of these QNMs $\Im \omega \propto \ln \sqrt{V_1 V_2}$, which $V_1, V_2$ denote the peak value of two separated potential respectively~\cite{Qian:2020cnz,Qian:2025occ}. In this paper, we focus on stability analysis in time domain, but our preliminary result of our future work in frequency domain shows that the imaginary part of \emph{fundamental mode} will change sign when $a \approx a_c$, implying that instability will always be triggered if the negative bump is sufficiently close to event horizon.

The ``negative bump'' we discussed here could destabilize the flat spacetime independently. However, the existence of BH will absorb energy and stabilize the spacetime. Thus, roughly speaking, the instability can be understood as the competition between the destabilizing caused by perturbed potential $V_p$ and the stabilizing caused by BH. One may expect that as long as the ``negative bump'' is sufficiently small and sufficiently close to the BH, the stable BH should dominate the stability of spacetime configuration, leading that entire spacetime configuration should tend to be stable. However, our results present a counterintuitive conclusion: no matter how small the bump is, it can overturn the stability of the entire spacetime, turning an originally stable Schwarzschild BH into an unstable one. Particularly, in the Appendix \ref{app-star}, we consider the situation that the background  is not a BH but a star, then we find that infinitesimal negative bump cannot trigger the instability if it is placed sufficiently close to the star. Thus, the instability discovered here not only is caused by negative bump but also involves intrinsic property of the black hole geometry.

\section{stochastic local potential}\label{slp}
In the second form we suppose $V_p(x,t)$ to be local stochastic function with zero mean. This in some sense is closer to real astrophysical environments compared with a fixed-sign perturbation. Many astrophysical BHs are surrounded by dynamical evolutional matter. The matter supplies non-static stochastic deformations on the geometry of BH. As a toy model, the effects of random matter then could be mimicked by random non-static modification on the Regge-Wheeler potential. More importantly, this kind of modification seemingly is inevitably even for an isolated BH without any matter surrounding it, since it can come from the local quantum fluctuations. Based on the equivalence principle, the wavelength $\lambda_p$ and period $t_p$ of local quantum fluctuation measured by local inertial observers are Planck scale $\lambda_p=t_p\sim\mathcal{O}(l_{\rm P})$. These quantum fluctuations will lead that the geometry of spacetime vary in the local \textit{proper time} scale $t_p$ and local \textit{proper distance} scale $\lambda_p$. Usually, such extreme ultraviolet variations of metric have no dramatic effect on classical physics. But the appearance of horizon makes things different due to the strong redshift.

To illustrate that, we consider the quantum fluctuation happening in the near-event-horizon region $r=r_h+\delta r$ (measured in coordinate gauge~\eqref{dsmetric}).  The line element in terms of tortoise coordinate in this region then will be $\td s^2 = h(r)(-\td t^2 + \td r_*^2) \approx  h'(r_h)\delta r (-\td t^2 + \td r_*^2)$ such that one finds $\lambda_p = \sqrt{h'(r_h) \delta r } \lambda =  \sqrt{h'(r_h) \delta r} T$. Here $\lambda$ and $T$ denote the wavelength and period in tortoise coordinate. The picture of classical horizon would fail if one is close to it near Planck scale. Thus, the proper distance between $r=r_h+\delta r$ and $r=r_h$ can at most as same as Planck scale. This means $\delta r /\sqrt{h(r)}\approx\delta r /\sqrt{h'(r_h) \delta r}\sim\mathcal{O}(\lambda_p)$. One thus finds that $T = \lambda  \sim 1/h'(r_h) \sim \mathcal{O}(r_h)$. Therefore, from the perspective of tortoise coordinate, near horizon local quantum fluctuations would induce metric or Regge-Wheeler potential to vary in infrared scale. It needs to emphasize again that, in the case of absent horizon or in the region far way from the horizon, the local quantum fluctuations will have no essential contributions to classical physics in infrared scale.

In order to characterize the randomness of variation on Regge-Wheeler potential, we specify
\begin{equation}\label{defwu1}
  V_p(x,t)= W(x)U(t)\,,
\end{equation}
where $W(x)$ is random function localizing around $x=0$ and $U(t)$ is a temporal random function. The function $W$ and $U$ are dimensionless functions and satisfy
\begin{equation}\label{eq1}
\overline{W}=\int^{\infty}_{-\infty} W(x) \td x=0,~\overline{U}=\lim_{t_m\rightarrow\infty}\frac1{t_m}\int^{t_m}_{-t_m} U(t) \td t = 0\,.
\end{equation}
The differences between $W$ and $U$ in Eq.~\eqref{eq1} are because that $W$ is required to localize around $x=0$ but function $U$ is assumed to have nonzero distribution in whole $t\in\mathbb{R}$. In addition, the spectrum (after of Fourier transformations) of $W$ and $U$ should mainly distribute in the region that are not larger than the order $\mathcal{O}(1/r_h)$.
\begin{figure}[hbpt]
  \centering
  \includegraphics[width=0.5\textwidth]{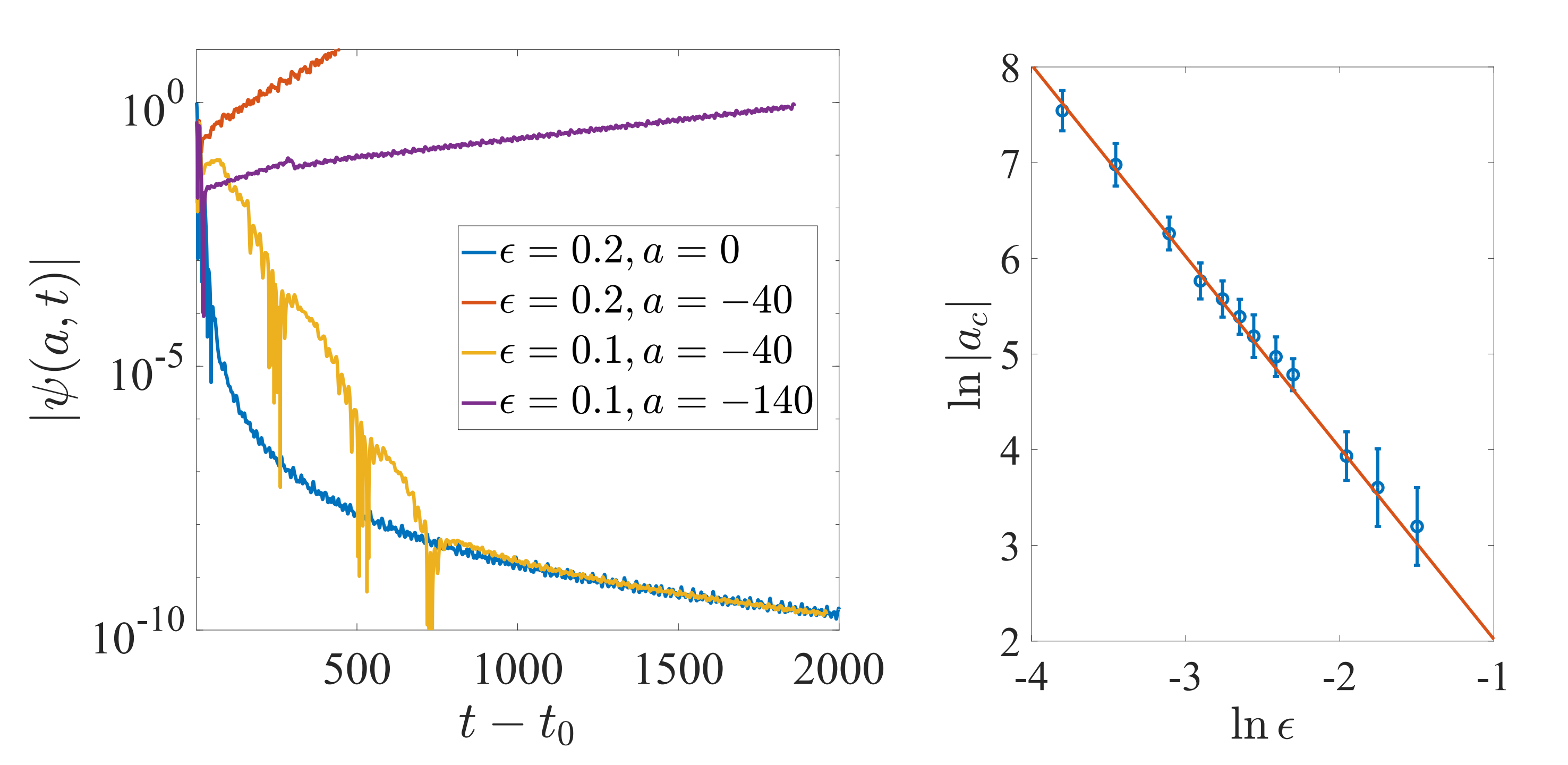}
  \caption{Left panel: Time evolutions of $\psi$ at $r_*=a$ for different values of $\epsilon$ and $a$. Right panel: Relationship between $\epsilon$ and $|a_c|$. The red line is $\ln|a_c|=-2\ln\epsilon+0.02$, which is obtained by fitting the most left 7 points. }\label{nonstaticfig1}
\end{figure}

As a concrete example, we choose pseudo-random functions without any explicit periodicity and symmetries such that $U(t)=\sin(t/2) + \cos (t/\sqrt{2}) - \sin(t/\pi)$ and
\begin{equation}\label{defex1}
W(x) = w_0\left[1-\tanh(4(x - 2)) \tanh(4(x+2))\right]Q(x)\,.
\end{equation}
with
\begin{equation}\label{defQ}
  Q(x)=\sin x + \cos \sqrt{2} x +\cos \pi x- \sin \pi x + c_0\,.
\end{equation}
Here $c_0$ and $w_0$ are chosen so that $W$ has zero average and normalization condition $\max|W|=1$ is satisfied. The detailed forms of Eq.~\eqref{defex1} are not important. We choose functions~\eqref{defex1} and \eqref{defQ} just as a concrete example. The results are shown in Fig.~\ref{nonstaticfig1}. We still find that there is a critical $a_c$ for every given amplitude $\epsilon$. The instability will be triggered if $a<a_c$.  The method of bisection gives asymptotical behavior
\begin{equation}\label{criticalac2}
  a_c\propto1/\epsilon^2
\end{equation}
for sufficiently small $\epsilon$. We have tested many different functions of $W$ and $U$ and find that Eq.~\eqref{criticalac2} is universal. The above numerical results still imply that the instability will be triggered if the random nonstatic deformation on the effective potential is sufficiently close to event horizon no matter how small the amplitude is.

\section{Low-frequency instability}
As we previously discussed, the stochastic local potential not only comes from the quantum fluctuations but also classical fluctuations. In Sec.~\ref{slp}, we focus on the quantum fluctuations with frequency near Planck scale, which will become $\mathcal{O}(1/r_h)$ due to the extreme redshift near BH horizon. Since the proper frequency of classical fluctuations are much lower than the quantum fluctuations, we thus consider the low frequency $w \ll {\cal O} (1/r_h)$ limit for the case of classical fluctuations near the BH horizons. We specify the perturbed potential $V_p$ to be
\begin{equation}\label{timedeVp1}
  V_p(r_*-a,t)=\exp\left[-\frac{(r_*-a)^2}{\sigma^2}\right][\sin(wt)+\sin(\pi wt)]\,,
\end{equation}
i.e, we consider a time-dependent perturbed potential that has Gaussian distribution in space but pseudo-random in time. Here we set $w\ll1$. In this situation, we will show that deformation potential itself will stabilize the flat background but can trigger the instability if it is closed to a BH.

\begin{figure}[hbpt]
  \centering
  \includegraphics[width=0.5\textwidth]{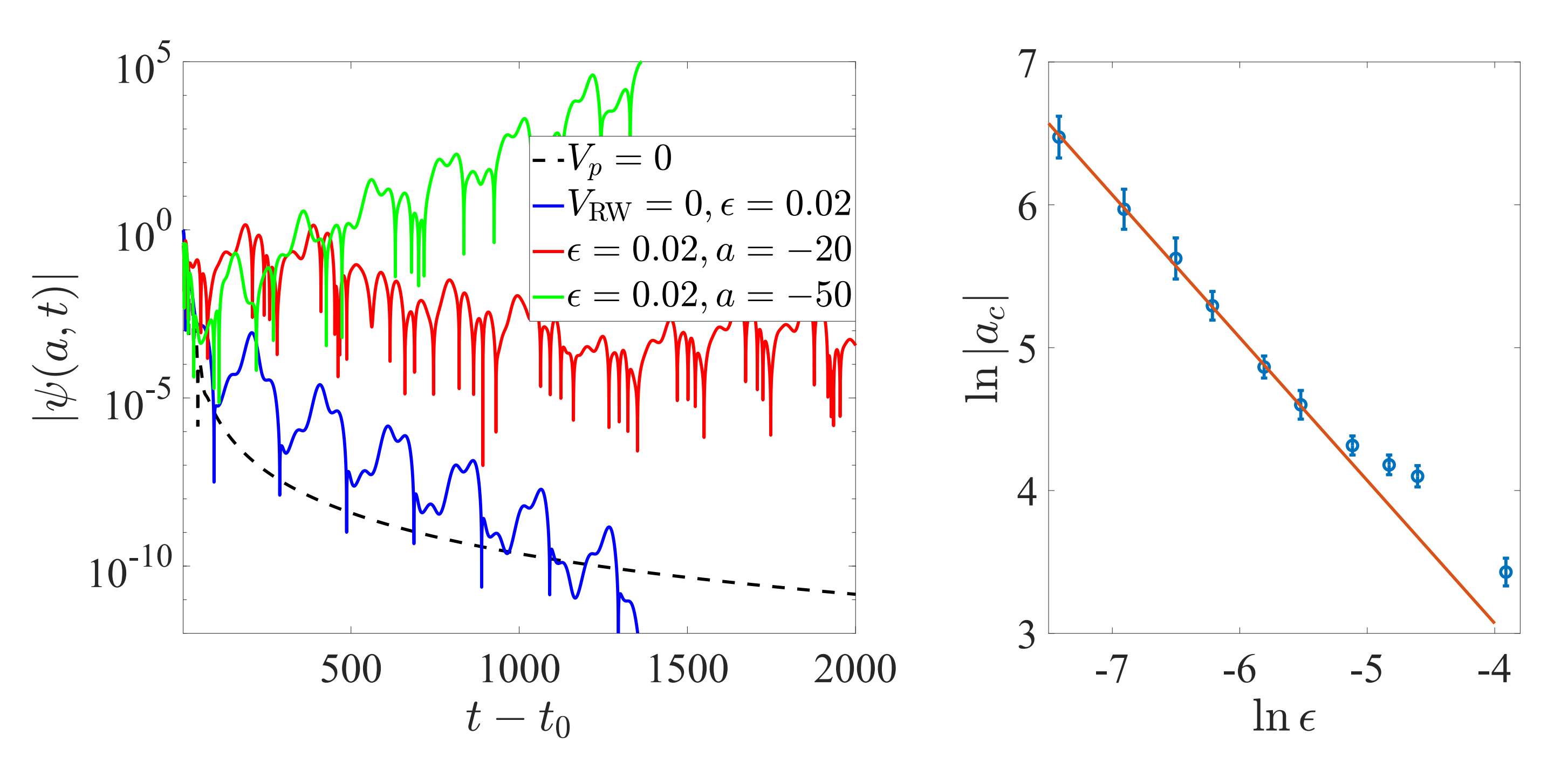}
  \caption{Left panel: Time evolutions of $\psi$ at $r_*=a$ in different situations with $w=0.02$ and $\sigma=4$. Right panel: Relationship between $\epsilon$ and $|a_c|$ for $w=0.001$ and $\sigma=4$. The red line is $\ln|a_c|=-\ln\epsilon-0.93$, which is obtained by fitting the most left 6 points. }\label{anomly1fig1}
\end{figure}
As one concrete example, we take $\sigma=4$ and $w=0.02$. The results are shown in Fig.~\ref{anomly1fig1}. If we turn off the BH potential (i.e. setting $V_{\text{RW}}=0$) but keep the perturbed potential $V_p$ by setting $\epsilon=0.02$, we find that the scalar probe field approximately exponentially decays into zero. Thus, the perturbed potential $V_p$ itself will stabilize the system. When we combine perturbed potential and Regge-Wheeler potential together, we find that there is critical distance $a_c\approx-35$ and the instability will be triggered if $a<a_c$. In the region $w\in[0.001,0.02]$ the results are similar. The numerical accuracy restricts us to explore smaller $w$. For sufficiently small $w$, saying $w=0.001$ for example, the method of bisection gives us following asymptotical behavior
\begin{equation}\label{criticalac3}
  a_c\propto-1/\epsilon
\end{equation}
for sufficiently small $\epsilon$. Our numerical results suggest that, for the case $w\rightarrow0^+$ the infinitesimal $\epsilon$ can still trigger the instability if the perturbed potential is sufficiently close to the horizon. This effect appears unique to BHs as these same potentials do not destabilize Minkowski spacetime.

\section{conclusion}
Astrophysical BHs are inevitably surrounded by matter distributions, and even isolated BHs experience quantum fluctuations. These classical and quantum perturbations induce stochastic deformations in spacetime geometry. This study addresses a critical question: whether nominally stable BHs retain their stability when the geometry is deformed by such small perturbations. It needs to emphasize that this question is not about the instability of a particular black hole but a question about the \emph{robustness of stability}.

To model this effect, we map the deformations onto modifications of the effective potential governing s-wave scalar fields. Numerical analysis provides compelling evidence that BH stability can be disrupted by infinitesimal negative or stochastic modifications localized near the event horizon. Notably, certain perturbed potentials that stabilize systems in flat backgrounds can destabilize otherwise stable BHs. While our focus is on the event horizon, analogous behavior may occur near cosmological horizons in asymptotically de Sitter BHs. Our findings suggest that the stability of a stable BH is not robust against the tiny deformation of near-horizon geometry. Infinitesimal deformation of near-horizon geometry can overturn the stability of a stable BH. Due to the inevitable fluctuations of near horizon geometry, this intrinsic instability fundamentally constrains our ability to predict BH dynamics over extended timescales.

As a toy model and very  preliminary research, we here only considered static spherically symmetric case and numerically simulated the system in time domain. It would be interesting to consider Kerr BH and study how spin would affect the phenomenon discovered here. Particularly, is there any relationship to turbulence of BHs discovered in Ref.~\cite{PhysRevLett.114.081101}?  Could we use frequency analysis to verify the results and supply deeper understandings on the underlying mechanism?~\footnote{The analysis of frequency domain for static negative bump and static stochastic potential will be addressed in an upcoming publication.} More realistically, could we quantitatively estimate the disturbance intensity $\epsilon$ generated by the actual environment (such as accretion disks, dark matter halos) or quantum fluctuations, and whether the position $a$ satisfies the condition of $a<a_c$? We hope these questions could be addressed in the future through numerical simulations and further theoretical work.

At the classical level, we emphasize that our results do not imply instability of the vacuum Schwarzschild solution, as no source exists to generate stochastic deformations on metric. For a BH surrounded by matter, our results imply that the system cannot keep static configuration for a \textit{long time} in astrophysical settings, since uncontrollable stochastic matter fluctuations near the horizon will lead to stochastic deformations on the geometry and inevitably trigger the instability identified here. The timescale that this instability becomes considerable depends on amplitude of the deformation. The probe scalar field studied here can act as a conduit, transferring energy from the environment to the BH and spatial infinity (See also Appendix \ref{energyflux}). Consequently, the final state evolves toward a Schwarzschild BH in vacuum, with increased horizon area and energy. However, if we posit an underlying quantum-gravitational framework
beneath classical general relativity, it becomes very interesting. We found that the quantum fluctuations could destabilize the Schwarzschild BH, implying that the scalar perturbations will exponentially grow with time and keep transferring energy to infinity. Will our finding be a potential mechanism for a new perspective on BH evaporation? This is a very interesting question. Further investigation is needed to establish a direct connection and we hope some useful advances will be obtained in the future.

%However, if we posit an underlying quantum-gravitational framework beneath classical general relativity, our analysis indicates that even vacuum Schwarzschild BHs become unstable. This suggests that Schwarzschild BHs in a quantum-gravitational framework would gradually lose energy over time, potentially offering a new perspective on BH evaporation. Further investigation is needed to establish a direct connection and we hope some useful advances will be obtained in the future.

\begin{acknowledgments}
We thank Prof.~Vitor Cardoso for sharing considerations on relationship between ``negative bump'' and instability of spacetime. We thank Yiqiu Yang for helpful discussions.  Z.~F.~Mai is supported by National Key R\&D Program of China (Grant Nos. 2024YFA1611704, and 2024YFA1611700). R.~Q.~Yang is supported by the National Natural Science Foundation of China under Grant No. 12375051. This work is also supported by the Guangxi Talent Program (``Highland of Innovation Talents'').
\end{acknowledgments}

\appendix
\section{Detailed Description of Numerical Integration Methods}\label{numer}
\begin{figure}[hbptp]
 centering
 \includegraphics[width=0.45\textwidth]{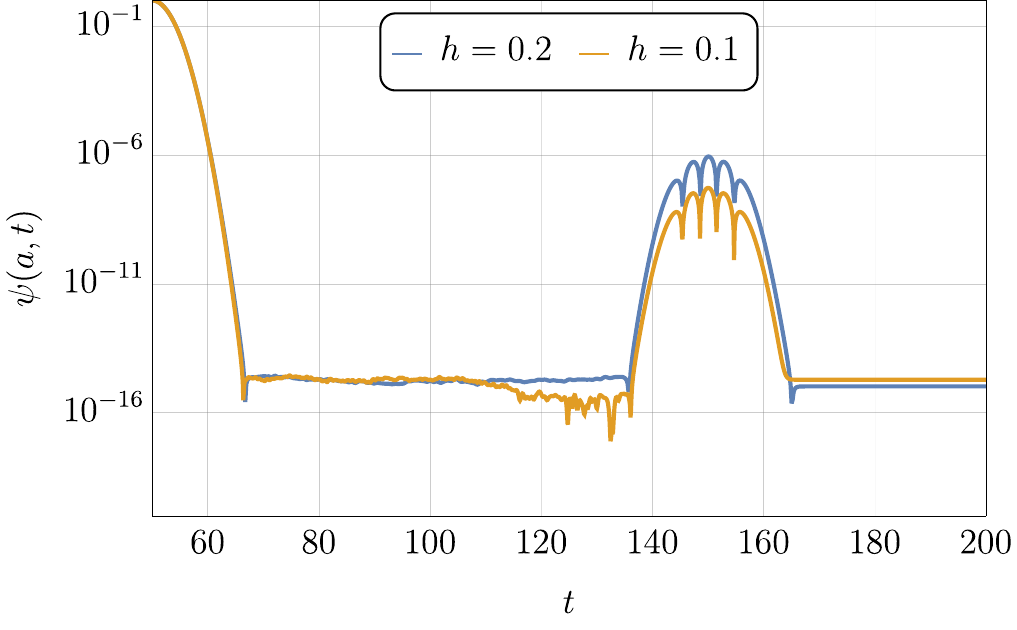}
 \caption{Time evolution of free scalar $|\psi(a,t)|$ at $a=-50$ without any potential using $h = 0.2$ and $h=0.1$.}\label{reflect}
\end{figure}

To obtain the finite difference equation of Eq.(2), we discretize the time coordinates $t = i \Delta t$ and the spatial coordinate $r_* = j \Delta r_*$, where $i= 0,1,2,\cdots N_{r_{*}}$ and $j = 0,1,\cdots N_t$ are integers. The scalar field $\psi$ and the effective potential $V_{\mathrm{eff}}$ are also discretized as
\begin{eqnarray}
&& \psi(t,r) = \psi_{i,j} = \psi(i\Delta t, j\Delta r_*),  ~\cr
~ \cr
&& V_{\mathrm{eff}} (r_*) = V_j \equiv V_{\mathrm{eff}} (j \Delta r_*) \, .
\end{eqnarray}
Eq.~(2) is first treated with a fourth-order centered finite difference scheme for the spatial second derivative $\frac{\partial ^2 \psi}{\partial r_*^2}$:
\begin{eqnarray}
\left. \frac{\partial ^2 \psi}{\partial r_*^2} \right|_{i,j} \equiv {\cal L}[\psi_{i,j}]&& \approx  \frac{1}{12 \Delta (r_*)^2} \left(-\psi_{i,j+2} + 16 \psi_{i,j+1}  \right.\cr
~\cr
&& \left.- 30 \psi_{i,j} +16\psi_{i,j-1} - \psi_{i,j-2}\right) \, .
\end{eqnarray}
To solve Eq.~(2) using fourth-order Rugge-Kutta method, we introduce an auxiliary variables $\phi_j = \frac{\partial \psi_j}{\partial t}$ and reduce it to a first-order system
\begin{equation}
\frac{\td \psi_j}{\td t} = \phi_j \, , \quad \frac{\td \phi_j}{\td t} = {\cal L}[\psi_j] - V_j \psi_j  \, .
\end{equation}
For each time step $i \to i+1$, we compute four intermediate Runge-Kutta stages as
\begin{eqnarray}
&& k_{1,\psi} = \Delta t \phi_{i,j} \, , \quad k_{1,\phi} = \Delta t ({\cal L}[\psi_{i,j}] - V_j \psi_{i,j}) \cr
~\cr
&& k_{2,\psi} = \Delta t \left(\phi_{i,j} + \frac{1}{2} k_{1,\phi}\right) \, , ~\cr
~\cr
&& k_{2,\phi} = \Delta t \left({\cal L}\left[\psi_{i,j} + \frac{1}{2} k_{1,\psi} \right]- V_j \left(\psi_{i,j} + \frac{1}{2} k_{1,\psi} \right)    \right) \, , ~\cr
~\cr
&& k_{3,\psi} = \Delta t \left(\phi_{i,j} + \frac{1}{2} k_{2,\phi}\right) \, , ~\cr
~\cr
&& k_{3,\phi} = \Delta t \left({\cal L}\left[\psi_{i,j} + \frac{1}{2} k_{2,\psi} \right]- V_j \left(\psi_{i,j} + \frac{1}{2} k_{2,\psi} \right)    \right)\, ,  ~\cr
~\cr
&& k_{4,\psi} = \Delta t \left(\phi_{i,j} + k_{3,\phi}\right) \, , ~\cr
~\cr
&& k_{4,\phi} = \Delta t \left({\cal L}\left[\psi_{i,j} + k_{3,\psi} \right]- V_j \left(\psi_{i,j} + k_{3,\psi} \right)    \right)
\end{eqnarray}
and update the solution as
\begin{eqnarray}
&& \psi_{i+1,j}  = \psi_{i,j} + \frac{1}{6} \left( k_{1,\psi} +  2k_{2,\psi}+ 2k_{3,\psi} +  k_{4,\psi}\right) ~\cr
~\cr
&& \phi_{i+1,j}  = \phi_{i,j} + \frac{1}{6} \left( k_{1,\phi} +  2k_{2,\phi}+ 2k_{3,\phi} +  k_{4,\phi}\right)  \, .
\end{eqnarray}
Furthermore,at the horizon ($r_*\rightarrow-\infty$) and the infinity  ($r_*\rightarrow\infty$) we impose the no reflecting boundary conditions. However, in practical numerical computations, $r_*$ cannot extend to $\pm\infty$, we therefore introduce a truncation at $r_* = \pm r_{*c}$. To insure the system can still simulate the situation of ``no boundary'' case, we impose the no reflecting boundary conditions as
\begin{equation}\label{findbd}
\left. \frac{\partial \psi}{\partial t}  -  \frac{\partial \psi}{\partial r_*} \right|_{r_* =- r_{*c} } = 0 \quad \left. \frac{\partial \psi}{\partial t}  +  \frac{\partial \psi}{\partial r_*} \right|_{r_*  = r_{*c}} = 0
\end{equation}
To ensure a fourth-order precision difference scheme, we introduce two additional auxiliary points at both boundaries respectively: $\psi_{-2,j}, \psi_{-1,j}, \psi_{N_{r_* + 1},j}, \psi_{N_{r_* + 2},j} $, satisfying
\begin{eqnarray}
\psi_{-1,j} = &  -5  \phi_{0,j} \Delta r_* + 10 \psi_{1,j} - 5 \psi_{2,j}   \cr
~\cr
& + \frac{5}{3} \psi_{3,j}- \frac{1}{4} \psi_{4,j} - \frac{65}{12}\psi_{0,j} \, , \cr
~\cr
\psi_{-2,j} = &  -30  \phi_{0,j} \Delta r_* + 80 \psi_{1,j} - 45 \psi_{2,j}  \cr
~\cr
& + 16 \psi_{3,j}- \frac{5}{2} \psi_{4,j} - \frac{95}{2}\psi_{0,j} \, ,  \cr
~\cr
\psi_{N_{r_* + 1},j} = &  -5  \phi_{N_{r_*},j} \Delta r_* + 10 \psi_{N_{r_*-1},j} - 5 \psi_{N_{r_*-2},j}  \cr
~\cr
& + \frac{5}{3} \psi_{N_{r_*-3},j}- \frac{1}{4} \psi_{N_{r_*-4},j} - \frac{65}{12}\psi_{N_{r_*},j} \, , \cr
~\cr
\psi_{N_{r_* + 2},j} = &  -30  \phi_{N_{r_*},j} \Delta r_* + 80 \psi_{N_{r_*-1},j} - 45 \psi_{N_{r_*-2},j}  \cr
~\cr
& + 16 \psi_{N_{r_*-3},j}- \frac{5}{2} \psi_{N_{r_*-4},j} - \frac{95}{2}\psi_{N_{r_*},j} \, . \cr
~\cr
\end{eqnarray}
In our practical calculation, we set $\Delta t = 0.9 \Delta r_{*}$, thus  for setting step size $\Delta r_* = h$, the accuracy of numerical results is ${\cal O} (h^4)$. To illustrate our numerical results on black hole instability are trustworthy, we show two results in Fig.~\ref{reflect} that time evolutions of free scalar field $\psi$ without any potential using $h = 0.2$ and $h=0.1$ (Corresponding to the blue solid line and orange solid line respectively). The ratio of two step sizes is $2$. The bump around $t=150$ is the reflecting wave from the left boundary. This reflection is due to our finite boundary effect and discretization error, which can be used as a character of accuracy for our numerical simulation. The maximum values of the reflecting wave are around $8.58 \times 10^{-7}$ for $h=0.2$ while around $5.08\times 10^{-8}$ for $h=0.1$. It is obviously that the ratio of these two  maximum values is approximately $16.8898$, which is close to $16$, the fourth power of the step size ratio. It shows that our numerical method indeed realizes desired convergency and so our numerical simulation is reasonable. In the main text, we typically set $h\sim0.01$ and so the reflection rate is about $\mathcal{O}(10^{-12})$. Such a reflection is negligible for the time scale considered here.

\section{Negative bump potential and dominant energy condition}\label{nebump}
In this appendix, we will give a concrete example that Regge-Wheeler potential has a local negative bump but dominant energy condition is satisfied.

Let us first begin from general static spherical asymptotically flat BH in tortoise coordinates, of which the metric reads
\begin{equation}\label{metrics1}
  \td s^2=h(x)(-\td t^2+\td x^2)+r(x)^2\td\Omega^2\,.
\end{equation}
We here assume that the event horizon locates at $x=-\infty$, where we have $r(-\infty)=r_h$ and $h(-\infty)=0$. The function $h(x)$ has asymptotical expression $h(x)=1-2M/x+\cdots$ with the total mass $M$ of the spacetime and  $r'(\infty)=1$.  The Regge-Wheeler potential $V_{\mathrm{RW}}$ for s-wave scalar field then reads
\begin{equation}\label{rwp1}
  V_{\mathrm{RW}}=r''/r\,.
\end{equation}
In static spherically symmetric case, the energy momentum tensor ${T^\mu}_\nu$ has a form ${T^\mu}_{\nu}=\text{diag}[-\rho(x),p_x(x),p_T(x), p_T(x)]$. The dominant energy condition then requires
\begin{equation}\label{dec1}
  \rho\geq |p_x|,~\rho\geq |p_T|\,.
\end{equation}
The Einstein's equation gives following independent equations.
\begin{eqnarray}
  h'&=&\frac{1+8\pi r^2p_x}{rr'}h^2-\frac{hr'}{r}\,,\label{einsth}\\
  rr''&=&h-r'^2-4\pi r^2(\rho-p_x)h\,\label{einstr}\,.
\end{eqnarray}
We combine Eq.~\eqref{einstr} and \eqref{rwp1} to obtain
\begin{equation}\label{rwp2}
  V_{\mathrm{RW}}=\frac{h-r'^2-4\pi r^2(\rho-p_x)h}{r^2}\,.
\end{equation}
For Schwarzschild BH, we have $\rho=p_x=0, h=1-2M/r$ and $r'=h$. We then recover the result $V_{\mathrm{RW}}=2M(1-2M/r)/r^3$.

We now consider the near horizon region in general case. From Eqs.~\eqref{einsth} and \eqref{einstr} one can find that there is no constraint on the values of $\rho-p_x$ and $p_x$. The values of $\rho$ and $p_x$ could be considerably large near horizon. Particularly, if an astrophysical BH is surrounded by sufficient matter localization around $x\sim a\ll -r_h$ and the dominant energy condition is satisfied, then it is possible
\begin{equation}\label{largematter}
  4\pi r^2(\rho-p_x)>1-r'^2/h\,
\end{equation}
and we then see that the sign of $V_{\mathrm{RW}}$ would be negative.

As a concrete example to confirm above statement, we take $p_x(x)=-\rho(x)$ and
\begin{equation}\label{derho}
  \rho(x)=\left\{\begin{split}
  b>0,~~~|x-a|<1/8\\
  0,~~~|x-a|>1/8
  \end{split}
  \right.\,
\end{equation}
as an example. To solve the Einstein equation, we note that the system is in vacuum and the solution is a Schwarzschild BH with Schwarzschild radius $r_h$ in the region $x<a-1/8$, of which the metric reads
\begin{equation}\label{valuexr}
  x=r+r_h\ln(1-r_h/r),~~r'=h=1-r_h/r\,.
\end{equation}
We then integrate equations~\eqref{einsth} and \eqref{einstr} into infinity with the initial values at $x=a-1/8$ given according to Eq.~\eqref{valuexr}.
\begin{figure}[hbpt]
 \centering
 \includegraphics[width=0.45\textwidth]{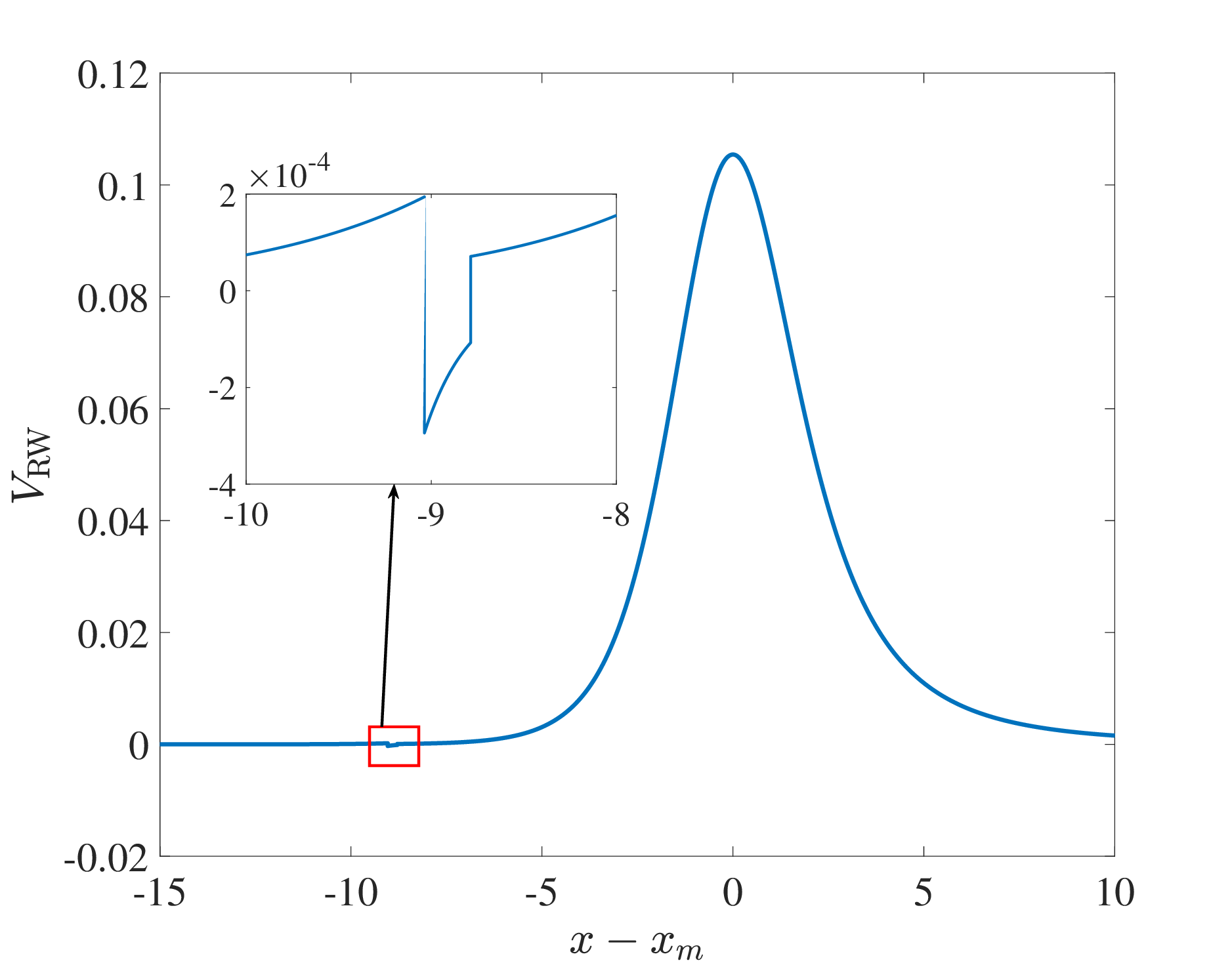}
 \caption{Regge-Wheeler potential for $a=-20,r_h=1$ and $b=0.2$. Here $x_m\approx-11.09$ is the position where $V_{\mathrm{RW}}$ reaches its maximum. }\label{negaVeff1}
\end{figure}
In Fig.~\ref{negaVeff1} we give a numerical result when we choose $a=-20, r_h=1$ and $b=0.2$. We see that there is negative bump around $x=a$. Thus, the negative bump potential can appear even if dominant energy condition is satisfied. We expect that, in general, localized spherical matter shells will introduce similar bumps in the Regge-Wheeler potential, possibly causing instabilities discussed in our main text.

\section{Infinitesimal negative bump in static spherical star}\label{app-star}
In this appendix, we consider a static ideal star as background, then we find that infinitesimal negative bump cannot trigger the instability if it is placed sufficiently close to the star. Since an astrophysical star involves much complicated details of its interior, we here just us a toy model and only regard the star as a horizonless geometry with nonzero total mass. We consider a regular Hayward star. The metric of the Hayward star is ~\cite{Hayward:2005gi}
\begin{equation}
\td s^2 = - h(r) \td t^2 + \frac{\td r^2}{h(r)} + r^2 \td \Omega^2\, , \quad h = 1-\frac{2M r^2}{r^3 + 2b^2M}\, .
\end{equation}
When $M < 3\sqrt{3}/4 b$, there is no horizon and the metric thus describe\ZF{s} a regular star. For without the loss of generality, we set that $M = 1, b=2$ as a concrete example. One important difference is that the tortoise coordinate $r_*$ here is defined as
\begin{equation}\label{defrstar1}
  r_*=\int_0^r\frac{\td x}{h(x)}\in[0,\infty)
\end{equation}
This leads fundamental changes: For BH, ``infinitely close to horizon'' means $r_*\rightarrow-\infty$; For a star, ``infinitely close to star'' means $r_*$ is still finite and can at most decrease to zero. This difference is universal for general spherically symmetric stars and BHs.
\begin{figure}[hbptp]
 \centering
 \includegraphics[width=0.45\textwidth]{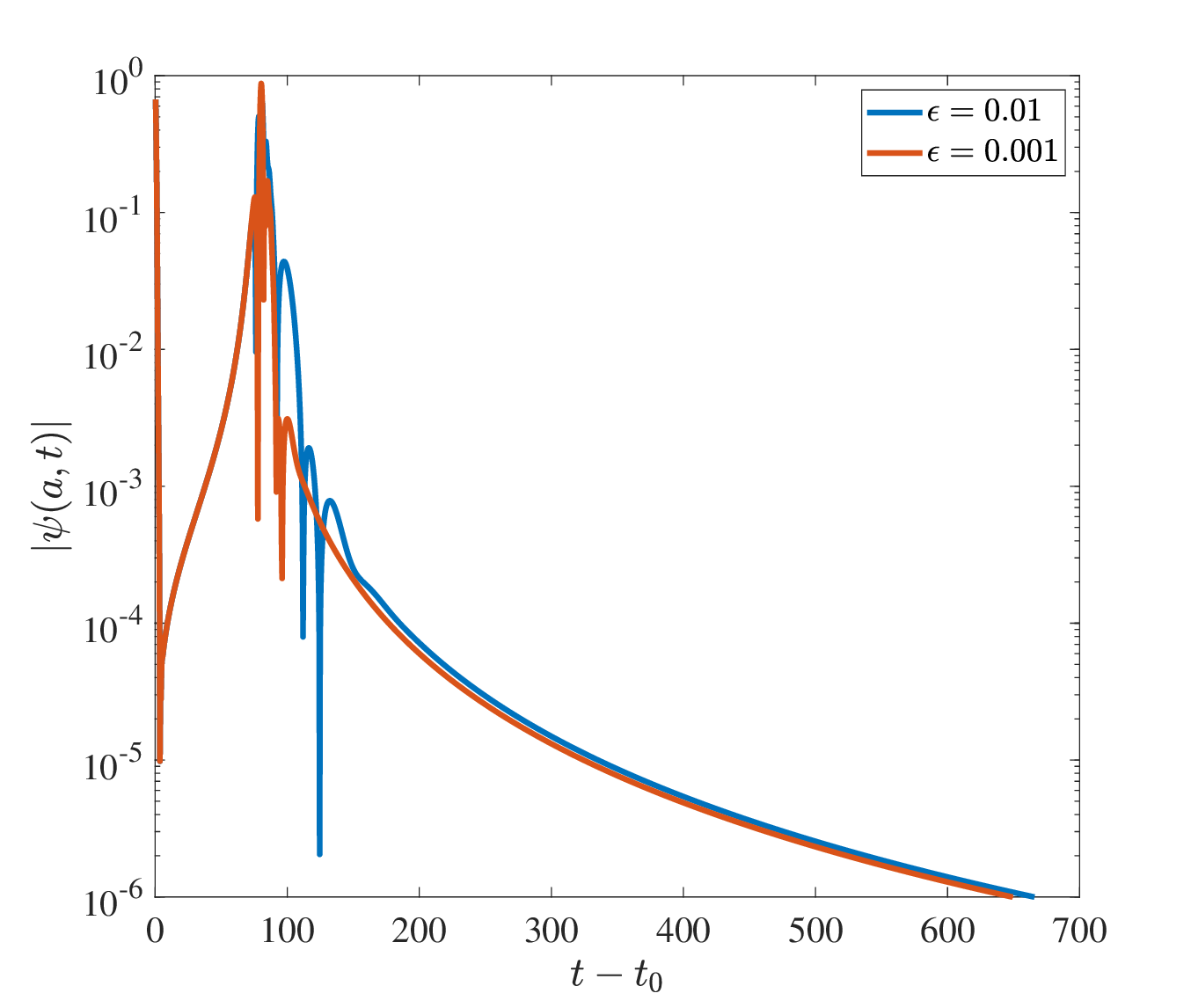}
  \caption{Time evolutions of $\psi$ at $r_*=a$ in different situations with $\epsilon=0.01$ and 0.001}\label{figstar}
\end{figure}

In Fig.~\ref{figstar}, we numerically show two cases of $a=0$: $\epsilon=0.01$ and $0.001$. We choose $a=0$ since this stands for the case that is closest to the star. Clearly, sufficiently small negative bump cannot trigger the instability when it is ``sufficiently close to the star''.\footnote{When the negative bump is sufficiently far way from the star, the instability can still be triggered but this is the effect of negative bump in flat space. }

\section{Energy flow of the scalar field}\label{energyflux}
To explain how the probe scalar field would serve as the conduit, we can compute the energy fluxes towards event horizon and infinity. For a light massless scalar field, the energy momentum tensor
\begin{equation}
T^{\mu\nu} = \nabla^\mu \Psi \nabla^\nu \Psi - \frac{1}{2}g^{\mu\nu} (\nabla \Psi)^2 \, ,
\end{equation}
where $g^{\mu\nu}$ denotes the metric of Schwarzschild black hole background and $\Psi=\psi/r$. It needs to note that asymptotical flatness requires $\Psi$ to be zero at infinity \textit{but} $\psi$ \textit{could be finite}. Since the background spacetime is assumed to be static with timelike Killing vector $\xi^\mu=(\partial/\partial t)^\mu$, then we can define the an energy current $J^\mu=-T^{\mu\nu}\xi_\nu$ and it satisfies $\nabla_\mu J^\mu=0$.

\begin{figure}[hbpt]
  \centering
  \includegraphics[width=0.28\textwidth]{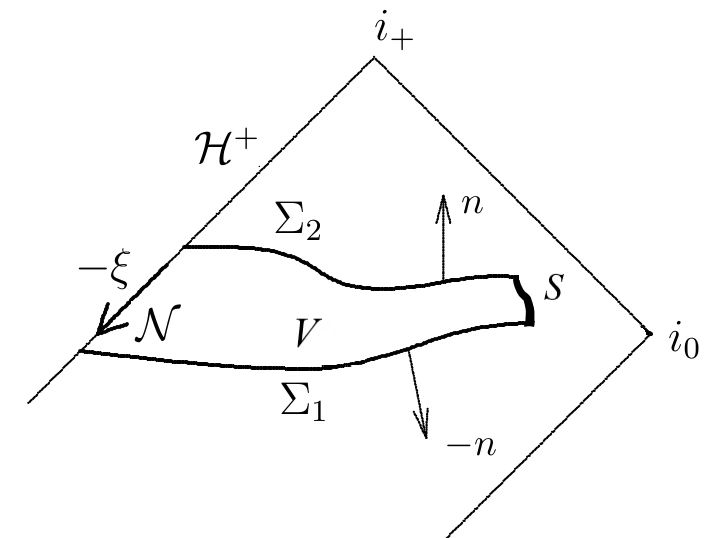}
  \caption{Penrose diagram on the exterior of horizon and a family of spacelike hypersurfaces $\{\Sigma_s\}$. Here $\mathcal{N}$ is boundary belong to horizon and $S$ is the boundary at $r=r_0\gg r_h$. In the limit $r_0\rightarrow\infty$, we have $S\subset i_0$. It should be note that $-\xi^\mu$ is `outward
   directed' normal to $\mathcal{N}$, as determined by continuity. }\label{sigma1}
\end{figure}

In order to study the energy fluxes at horizon and infinity, let us consider a family of spacelike hypersurfaces $\{\Sigma_s\}$ outside event horizon, which gives a foliation on the spacetime region between $r_h$ and $r=r_0\gg r_h$. We assume that this foliation preserve the spherical symmetry. The boundaries of every $\Sigma_s$ are given by $r=r_h$ and $r=r_0$. The energy in one hypersurface $\Sigma_s$ is given by
\begin{equation}\label{defEs}
  E_s=\int_{\Sigma_s}J^\mu\td\Sigma_\mu\,.
\end{equation}
Here $\td\Sigma_\mu$ is the future-toward directed volume element. Let us consider two different hypersurface $\Sigma_1$ and $\Sigma_2$, see Fig.~\ref{sigma1}.
The difference of energies between these two hypersurfaces is
\begin{equation}\label{defEs1}
\begin{split}
  &E_2-E_1=\int_{\Sigma_2}J^\mu\td\Sigma_\mu-\int_{\Sigma_1}J^\mu\td\Sigma_\mu\\
  &=\oint_{\partial V}J^\mu\td\Sigma_\mu-\left(\int_{\mathcal{N}}J^\mu\td\Sigma_\mu+\int_{S}J^\mu\td\Sigma_\mu\right)\,.
  \end{split}
\end{equation}
Here $\mathcal{N}$ is boundary belong to horizon, $S$ is the boundary at $r=r_0$ and $V$ is the spacetime region surrounded by $\Sigma_1, \Sigma_2, \mathcal{N}$ and $S$. The hypersurface $S$ is define by $r=r_0$ and $t\in[t_1,t_2]$. The null hypersurface $\mathcal{N}$ is defined by $r=r_h$ and $v\in[v_1,v_2]$, where $(\partial/\partial v)^\mu$ is the future-directed generator of event horizon. At event horizon, $\xi^\mu$ is both normal and tangent to event horizon and we take $(\partial/\partial v)^\mu=\xi^\mu$.

After using Gaussian theorem and $\nabla_\mu J^\mu=0$, we  then have
\begin{equation}\label{defEs2}
  E_2-E_1=-\left(\int_{\mathcal{N}}J^\mu\td\Sigma_\mu+\int_{S}J^\mu\td\Sigma_\mu\right)\,.
\end{equation}
The above two integrations describe the energy fluxes at horizon and $r=r_0$, respectively. Since $r=r_0$ is a timelike surface and its unit outer normal vector reads $n^\mu=(\partial/\partial r)^\mu$ (we assume that $r_0\gg r_h$). The energy flux at $r=r_0$ then is given by
\begin{equation}\label{energyS1}
\begin{split}
  \int_{S}J^\mu\td\Sigma_\mu&=-4\pi r_0^2\int_{t_1}^{t_2}\td t\dot{\Psi}\partial_r\Psi|_{r=r_0}\\
  &=\left.-4\pi \int_{t_1}^{t_2}\td t\dot{\psi}\left(\frac{\partial\psi}{\partial r}-r^{-1}\psi\right)\right|_{r=r_0}\,.
  \end{split}
\end{equation}
The ``no-reflect boundary condition'' at the infinity shows that
$$\left.\dot{\psi}+\frac{\partial\psi}{\partial r}\right|_{r\rightarrow\infty}=0\,,$$
and asymptotical flatness requires $\psi/r=\Psi\rightarrow0$. The energy flux at $S$ then is positive and is toward infinity. Since $\mathcal{N}$ is a null surface and its outer normal vector reads $-\xi^\mu$ (we assume that $r_0\gg r_h$). The energy flux at $\mathcal{N}$ then is given by
\begin{equation}\label{energyS2}
\begin{split}
  &\int_{\mathcal{N}}J^\mu\td\Sigma_\mu=-\int_{\mathcal{N}}J^\mu\xi_\mu\td\Sigma\\
  &=4\pi r_h^2\int_{v_1}^{v_2}\td t\dot{\Psi}\partial_t\Psi|_{r=r_0}=4\pi\int_{v_1}^{v_2}\td v\dot{\psi}^2\,.
  \end{split}
\end{equation}
Here we have used the fact that $\xi^\mu$ is null vector at horizon. The energy flux towards interior of event horizon then is also positive.

In the stable case, any fluctuation of test scalar field will decay to zero fast, so energy fluxes at $\mathcal{N}$ and $S$ will also decay to zero fast. However, in the situations discussed in main text, the fluctuation of test scalar field will exponentially increase, so energy fluxes at $\mathcal{N}$ and $S$ will be more and more. Therefore the scalar field acts as a conduit, transferring energy from the environment into the black hole and out to infinity.

\begin{figure*}[!t]
\centering
\includegraphics[width=\textwidth]{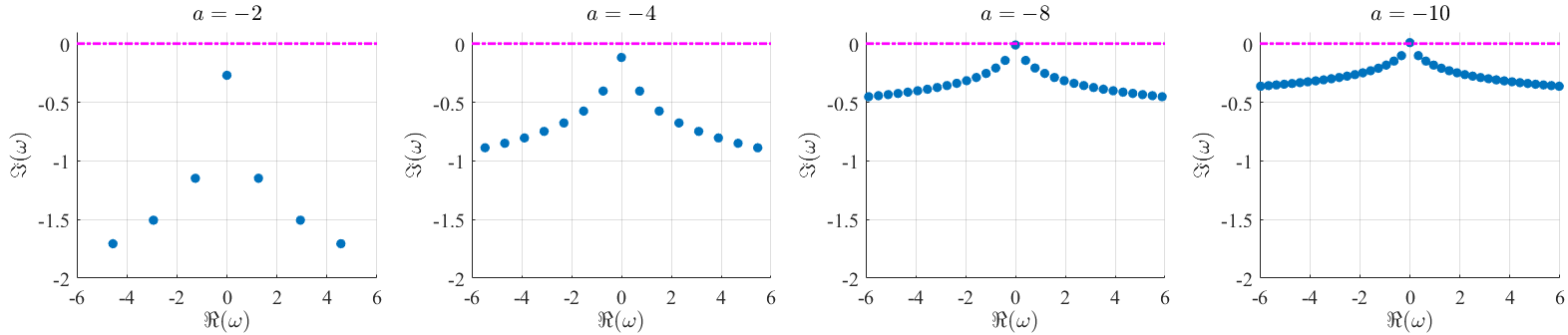}
\caption{QNMs of potential~\eqref{deltap0} for different distances. Here we set $V=1$ and $\epsilon=0.1$. The critical distance $a_c\approx-10$ and fundamental mode raises into upper complex plane when $a<a_c$. }\label{deltap1}
\end{figure*}
\section{$\delta$-function potential}
In this appendix, we consider a special type of effective potential, namely the $\delta$-function type potential to illustrate the instability semi-analytically from the frequency domain. We will use this us a tool model to understand why the instability happen in the frequency domain. Furthermore, we shall reproduce the asymptotic relation between the $\epsilon$ and $a_c$. We begin with the equation of a scalar field $\Psi$
\begin{equation}\label{maseq2}
\frac{\partial^2 \Phi}{\partial t^2} - \frac{\partial ^2 \Phi}{\partial x^2} + V \Phi = 0\, , \quad x\in(-\infty,\infty)
\end{equation}
We shall analyse the instability in frequency domain. Consider a Fourier transformation $\Phi(t,x) = \int \td \omega  e^{-\ti\omega t} \phi (\omega, x)$, we have
\begin{equation}\label{maseq3}
\frac{\td^2 \phi}{\td x^2}+(\omega^2 - V) \phi = 0\, , \quad x\in(-\infty,\infty)
\end{equation}
To analyse the (in)stability of  the scalar field, we impose the following boundary condition as
\begin{eqnarray}\label{sol2}
    \left.\phi \right|_{x \to -\infty} = \te^{- \ti \omega x} \,  , \quad \left.\phi \right|_{x \to \infty} = \te^{\ti \omega x} \, .
\end{eqnarray}
Such boundary condition will select a series discrete complex spectrum, namely the QNMs of the scalar field. Furthermore, the QNMs of the scalar filed can be a indicator of the instability. The positive/negative imaginary part of the QNMs implies unstable/stable scalar field.

Without loss of generality, we consider the following two delta function potential
\begin{equation}\label{deltap0}
V(x) = -\epsilon \delta (x - a) + V \delta(x)\, , \quad {\epsilon} >0 \quad V >0
\end{equation}
implying that two delta function type potential localized at $x=a$ and $x=0$ respectively. Here the ${V}$ terms is used to analogize the black hole potential and the ${\epsilon}$ term is used to analogize the negative bump. Moreover, $\delta(x - x_0)$ denotes the Dirac delta function, satisfying that
\begin{eqnarray}
\delta(x - x_0) =
\begin{cases}
0 \quad & x \neq x_0 \\
\infty \quad & x = x_0
\end{cases}
~~ \int^\infty_{-\infty} \delta (x-x_0) \td x = 1 \, .~\cr
\end{eqnarray}
Based on the property of delta function potential, $\phi$ is continuous but its derivative $\phi'$ is discontinuous at $x = 0$ and $x = a$. Considering the boundary condition of the QNMs, $\phi(x)$ will be divided into three part (assume $a<0$)
\begin{equation}
\phi(x) =
\begin{cases}
\phi_{\mathrm{I}} = \te^{-\ti \omega x}  \quad     & -  \infty < x < a \\
\phi_{\mathrm{II}} = A \te^{\ti \omega x} + B \te^{-\ti \omega x}  \quad    &  a< x <0 \\
\phi_{\mathrm{III}} = C \te^{\ti \omega x}  \quad   & 0 < x < \infty \\
\end{cases} \, .
\end{equation}
The joint conditions read
\begin{eqnarray}
&&\begin{cases}
\phi_{\mathrm{I}} (a) = \phi_{\mathrm{II}} (a) \\
\phi'_{\mathrm{II}} (a) - \phi'_{\mathrm{I}} (a) = -{\epsilon} \phi(a)
\end{cases}
~\cr
~\cr
&&\begin{cases}
\phi_{\mathrm{II}} (0) = \phi_{\mathrm{III}} (0) \\
\phi'_{\mathrm{III}} (0) - \phi'_{\mathrm{II}} (0) =  {V} \phi(0)
\end{cases} \, .
\end{eqnarray}

These four joint conditions can eliminate the coefficients $A,B,C$ and give the equation for QNMs
\begin{equation}\label{deqnm}
\te^{-\ti a \omega} {\epsilon} {V} - \te^{ \ti a \omega} ({\epsilon} + 2 \ti \omega) ({V} - 2 \ti \omega) = 0 \, .
\end{equation}
If setting ${\epsilon}=0$, i.e., there is only one positive potential, one then obtain
\begin{equation}
{V} - 2 \ti \omega = 0 \, ,
\end{equation}
and find that there exists only one QNM,
\begin{equation}\label{deqnm2}
\omega_{\text{QNM}} =- \frac{\ti}{2} {V} \, .
\end{equation}
It shows  that the QNM of the delta potential is a pure decay mode. If setting ${\epsilon} \neq 0$ and $a\neq0$, which implies that the ``negative bump'' ${\epsilon}$ is open. Eq.~\eqref{deqnm} has no analytical solution but numerical solution. We give the Fig.~\ref{deltap1} showing the relation of the imaginary part of the QNM and $a$. Though the $\delta$-function is seemingly to ``toy'' model comparing with the potentials studied in this paper, it clearly uncovers two main features that found by time-domain analyses in the main text, which implies the university of our founding:
\begin{enumerate}
\item[(1)] Competition between the destabilizing caused by negative bump and stabilizing caused by positive bump. From Eq.~\eqref{deqnm2} we see that the positive $\delta$-potential gives QNMs under the bottom complex plane and so will stabilizes the system; negative $\delta$-function gives QNMs under the upper complex plane and so will destabilizes the system. When small negative $\delta$-potential appears nearby a large positive $\delta$-potential, we see that system is still stable.
\item[(2)] Existence of critical distance. If their distance is large enough, the fundamental mode of QNM goes to upper complex plane and so the the system becomes unstable. The critical distance $a_c\approx1/\epsilon$, just as we have found the time-domain analyses.
\end{enumerate}

%The Fig.[xxx] shows that there exist a critical distance $a_c$ such that the imaginary part of QNM changes sign, indicating the stable mode changes to unstable one.
Furthermore, based on  Eq.~\eqref{deqnm}, we can also obtain a relation between $a_c$ and ${\epsilon}$ similar to that for Eq.~(6). Near the turning point of the QNM, the fundamental mode $\omega_0$ will be small and so we can expand  Eq.~\eqref{deqnm} up to  ${\cal O}(\omega^2)$ and obtain the critical distance $a_c = -\frac{{V} - {\epsilon}}{{\epsilon} {V}} $. If assuming that $\epsilon \ll {V}$, indicating that the amplitude of ``negative bump'' is small enough, we then find $a_c \approx -1/\epsilon$. This satisfies Eq.~(6) since $\bar{V}_ p = -\epsilon\int^\infty_{-\infty} \delta (x + a) \td x = -\epsilon $.

Besides, we also find that recently G. R. Li et al. also considered two delta-function type potential as a toy model to study a mapping between bound states and black hole QNMs \cite{Li:2026ptw}. Unlike our consideration of the variation in fundamental mode with $a$, they considered the asymptotic echo modes, namely $\omega_n \to \infty$, finding that the imaginary part of these modes $\Im \omega_n \propto \ln \sqrt{\epsilon V}$.

\bibliography{unBH_ref}
\end{document}